\documentclass[a4wide,twocolumn,prb,showpacs,amsmath,amssymb]{revtex4-1}

\usepackage{dcolumn}
\usepackage{bm}
\usepackage{epsf}
\usepackage{multirow}
\usepackage{graphicx}
\usepackage{amsmath}
\usepackage{color}
\usepackage{newunicodechar}
\usepackage{mathtools}
\usepackage{tabularx,ragged2e,booktabs,caption}

\begin{document}
\preprint{APS/123-QED}

\title{Raman energy density (RED) in the context of acousto-plasmonics}

\author{Jos\'{e} Luis Monta\~{n}o-Priede}
  \affiliation{Department of Physics and Astronomy, The University of Texas at San Antonio, One UTSA circle, San Antonio, Texas 78249, United States}
  \affiliation{Current address: Department of Electricity and Electronics, FCT-ZTF, UPV-EHU, Bº Sarriena s/n, 48940, Leioa, Bizkaia, Spain}

\author{Adnen Mlayah}
  \affiliation{Centre d'Elaboration de Mat\'{e}riaux et d'Etudes Structurales (CEMES), CNRS and Universit\'{e} Paul Sabatier - Toulouse III, 29 rue Jeanne Marvig, BP 94347, 31055 Toulouse cedex 4, France}
  \affiliation{Laboratoire d'Analyse et d'Architecture des Syst\`{e}mes (LAAS), CNRS and Universit\'{e} Paul Sabatier - Toulouse III, 7 avenue du Colonel Roche, BP 54200, 31031 Toulouse cedex 4, France}

\author{Nicolas Large}
  \email{nicolas.large@utsa.edu}
  \affiliation{Department of Physics and Astronomy, The University of Texas at San Antonio, One UTSA circle, San Antonio, Texas 78249, United States}
  

\date{\today}

\begin{abstract}
Interactions between elementary excitations are of great interest from a fundamental aspect and for novel applications. While plasmon-exciton have been extensively studied, the interaction mechanisms between acoustic vibrations (phonons) and localized surface plasmons (LSPs) remain quite unexplored. Here, we present a new theoretical framework for the investigation of the interaction between confined acoustic vibrations and LSPs involved in resonant acoustic Raman scattering. We express the Raman scattering process in the framework of Fermi's golden rule and introduce the concept of Raman energy density (RED). Similarly to the Raman-Brillouin electronic density (RBED) introduced for semiconductors, this physical quantity is used as a theoretical tool for the interpretation of resonant Raman scattering mediated by LSPs in metallic nanoparticles. The RED represents the electromagnetic energy density excited in the Raman scattering process and modulated by the acoustic vibrations of the nanoparticle. We show that, similarly to the LDOS and the RBED, the RED can be mapped in the near-field region, and provides a clear picture of the interaction between LSPs and acoustic vibrations giving rise to inelastic scattering measurable in the far-field. We use the newly introduced RED concept to investigate elastic (an)isotropy effects and extract the Raman selection rules of spherical nanoparticles embedded in a dielectric environment.
\end{abstract}


\maketitle

\section{Introduction}
Owing to their properties, which are governed by their size, composition, shape, environment, and interaction with neighboring nano-objects,\cite{Link1999, Kelly2003, Grady2004, Gonzalez2007, Montano2016, Martinsson2016} localized surface plasmons (LSPs) have proven to be very valuable for the realization of many applications.\cite{Large2010a, Doiron2019, Ozbay2006} This past decade, a great deal of effort has been made to improve the performance of these plasmonic-based applications. One strategy has been to couple LSPs to other elementary excitations, including photonic modes,\cite{Park2015, Wang2015} magnetic modes,\cite{Sachan2014, Ge2015} and excitons.\cite{Schlather2013, Abid2018} To date, this strategy has been very successful as it led to the emergence of new fundamental concepts\cite{Fofang2008, Sachan2014} and technological applications.\cite{Zhou2013} Most of the studies focusing on metallic nanoparticles (NP) consider them as static bodies at rest. However, LSPs can be temporally modulated by acoustic vibrations, also known as acoustic phonons, naturally present in the nanomaterial and surrounding environment. Very recently, there has been a renew of interest on the modulation of the LSPs by (opto)mechanical modes and elastic waves.\cite{Juodenas2020, Oumekloul2020, Cunha2020, SaisonFrancioso2020, Apell2022, Deacon2017} Such high frequency (GHz-THz) modulation can be used to increase detection and sensing accuracy of nano-objects and molecules. For instance, these vibrating nanoparticles, acting as simple optomechanical nanoresonators, can be seen as nanoscale analogs to quartz crystal microbalances.\cite{Girard2016a} Another work by Yoo \textit{et al.} showed that epsilon-near-zero (ENZ) cavities can be used to achieve ultra-strong coupling between ENZ gap plasmon modes and optical phonons.\cite{Yoo2021} 

  Acoustic phonons Raman scattering is a very effective, high precision, and non-invasive technique for nanometrology. It uses the acoustic vibrations as local probes to determine sizes and distances at the nanoscale with very high accuracy.\cite{Cazayous2001, Cazayous2002} However, the effectiveness of this technique is directly related to the knowledge of the interaction mechanisms between electrons (plasmons) and phonons (acoustic vibrations). Therefore, it is necessary to know what are the vibrational modes sustained by a given nanostructure (amplitude, frequency, symmetry) and know how they interact with the LSPs to completely determine the Raman efficiencies and selection rules.\cite{Duval1992}
  
  Here, we investigate the dynamic properties of metallic nanoparticles by focusing on the interaction between confined acoustic vibrations and localized surface plasmons.\cite{Large2009} The dynamic properties of LSPs are responsible for the transient optical absorption modulation, that can be observed in time-resolved transient absorption experiments,\cite{Marty2011, OBrien2014, Kirschner2016, Lethiec2016, Yi2017, Ahmed2017, Ahmed2022, Deacon2017} and for the acoustic SERS effect.\cite{Tripathy2011, Girard2016, Girard2017, Apell2022} The study of the acousto-plasmonic properties gives unprecedented insight into the fundamental interactions between elementary excitations at the nanoscale and opens up new ways to probe the mechanical properties of nanostructures using optical spectroscopy.\cite{Marty2013} Particularly, the acousto-plasmonic interaction in metallic nanoparticles has been studied in low-frequency Raman scattering experiments, in the 2 to 50 cm$^{-1}$ spectral region, where light is scattered by acoustic vibrations in the nanostructure.\cite{Tripathy2011} Acousto-plasmonic-driven Raman scattering can be described in three steps: i) incident photon absorption, ii) phonon emission/absorption by the lattice, and iii) scattered photon emission.\cite{Bachelier2004, Large2009} The photon-phonon interaction is not direct, but rather occurs \textit{via} the acousto-plasmonic coupling, \textit{i.e.}, photon-electron-phonon interaction, which results in the direct modulation of the LSP by the acoustic vibration.\cite{Large2009} When the metal nanostructure is optically excited close to its LSP resonance (LSPR), it leads to a strong acousto-plasmonic interaction.\cite{Bachelier2007} Similarly to the concept of near-electric-field hot-spots in plasmonic nanostructures, the sites where acoustic vibrations produce large modulations of the plasmon-induced localized electric field at the nanostructure surface are called acousto-plasmonic hot-spots.\cite{Large2009}
  
  In this work, we described the resonant Raman scattering process using a new approach based on a single direct acousto-plasmonic interaction process described by Fermi's golden rule. Within this framework, we introduce a new physical quantity, namely the Raman energy density (RED). Similarly to the Raman-Brillouin electronic density (RBED) introduced in semiconducting nanostructures,\cite{Huntzinger2006, Mlayah2007, Large2010} the RED allows for studying and monitoring the acousto-plasmonic Raman scattering in the near-field. We modeled the acousto-plasmonic interaction by implementing vibrational dynamics (resonant ultrasound method, RUS),\cite{Visscher1991} into electrodynamic calculations (discrete dipole approximation, DDA).\cite{Draine1994} We used this methodology to compute the acoustic Raman spectra and investigate the interaction between LSPs and (an)isotropic acoustic vibrations, thus leading to the full determination of the Raman efficiencies and selection rules. We show that the RED, which can be mapped in the near-field for each vibration mode, correlates the far-field Raman scattering to the local acousto-plasmonic hot-spots, provides a deeper understanding of the Raman scattering process (\textit{e.g.}, Raman selection rules), and serves to study the acousto-plasmonic coupling of complex nanostructures.\cite{Temnov2012, Mrabti2016, Yi2017}

\section{Theoretical Formalism}
\subsection{Raman Energy Density (RED)}
Fermi's golden rule gives the rate at which transitions take place between an initial and a final state of the system consisting of interacting plasmons and phonons in our case.\cite{Feldman1986} In this framework, the Raman scattering process is described as a single step transition between plasmon and phonon states described in the occupation number representation by $\lvert {n_{\rm pl}, n_{\rm ph}}\rangle$ ($n_{\rm pl}$ and $n_{\rm ph}$ being the occupation of the plasmon and phonon states). The probability of transition per unit time associated with the Raman scattering process is expressed as
\begin{equation}\label{Fermi}
\mathcal{R}^{(1)}_{fi}=\frac{2\pi}{\hbar}
\left|\langle n_{{\rm pl},f}, n_{{\rm ph},f} \lvert\hat{H}_{\rm{int}}\rvert n_{{\rm pl},i}, n_{{\rm ph},i}\rangle\right|^2 \delta(E_f-E_i),
\end{equation}
where $E_{i(f)}$ is the energy of the initial (final) plasmon-phonon state. For phonon emission (Stokes Raman scattering) $E_f-E_i = \hbar\omega_{{\rm pl},f}+\hbar\omega_{{\rm ph}}-\hbar\omega_{{\rm pl},i}$, where $\omega_{{\rm pl}}$ and $\omega_{{\rm ph}}$ are the plasmon and phonon frequencies. $\hat{H}_{\rm{int}}$ is the Hamiltonian for the acousto-plasmonic interaction and is expressed in the formalism of the second quantization as
\begin{equation}\label{hamiltonian}
\hat{H}_{\rm int}=\sum_{k}\int_{\rm V}{-E^*_{s,k} \frac{\partial P_{i,k}}{\partial k} u^*_{m,k}dV} ~ \hat{a}_s^{\dagger}\hat{a}_i\hat{b}_m^{\dagger},
\end{equation} 
where $\hat{a}^{\dagger}$ (resp. $\hat{a}$) and $\hat{b}^{\dagger}$ (resp. $\hat{b}$) are the plasmon and phonon creation (resp. annihilation) operators and $E_{s,k}$, $P_{i,k}$, and $u_{m,k}$ are the $k$-component of the classical electric field, polarization, and displacement vector, respectively.

  In previous works,\cite{Large2009, Bachelier2004} the Raman scattering process was described using a third-order perturbation theory approach. In this three-step process, the surface plasmon is excited by the incoming photon and decays into another surface plasmon state \textit{via} emission or absorption of confined vibration, and finally a scattered photon is emitted by this second excited plasmon state. However, it is important to notice that our approach use Fermi’s golden rule to describe a one-step transition between two excited plasmon states because the plasmonic fields calculated though electrodynamic simulations already account for the optical excitation and emission processes (see Supplementary Materials\cite{SupMat}).
  
  In the absence of any vibration (\textit{i.e.}, nanoparticle at rest), no transition can occur since the LSP states form a set of orthogonal eigenstates. When the nanoparticle vibrates, the LSP polarization vector is modulated, thus enabling dipolar transitions between LSP states.\cite{Bachelier2004} The interaction matrix element between the confined acoustic vibrations and the LSP states is expressed for phonon emission as\cite{Large2009}
\begin{equation}\label{interaction1}
  \langle 1_{f}, 1_{m} \left|\hat{H}_{\rm{int}}\right|1_{i}, 0\rangle = -\int \mathbf{E}_f(\mathbf{r})\cdot\delta_{m}\mathbf{P}_i(\mathbf{r})dV,
\end{equation} 
where $\mathbf{P}_i(\mathbf{r})$ is the LSP-induced polarization when excited in the Raman scattering process (initial LSP state) and modulated by the confined acoustic vibration mode $m$ ($\delta_{m}\mathbf{P}_i$) and $\mathbf{E}_f(\mathbf{r})$ is the local electric field associated with the final LSP state, which gives rise to the scattered light experimentally detected. Using this interaction matrix element and defining the Raman energy density (RED) as $\mathcal{U}_{\text{R},m}(\mathbf{r},\omega_{i},\mathbf{k}_i,\omega_{f},\mathbf{k}_f)\equiv-\mathbf{E}_f(\mathbf{r})\cdot\delta_{m}\mathbf{P}_i(\mathbf{r})$ we can express the transition rate (Eq.~\ref{Fermi}) as
\begin{equation}\label{Fermi_RED}
  \mathcal{R}^{(1)}_{fi}=\frac{2\pi}{\hbar}\left| \int \mathcal{U}_{\text{R},m}(\mathbf{r},\omega_{i},\mathbf{k}_i,\omega_{f},\mathbf{k}_f) dV \right|^2 \delta(E_f-E_i).
\end{equation} 
The RED is a complex local energy density that gives rise to Raman scattering. It contains the excitation, interaction, and emission steps of the Raman scattering process and is in $\rm{J/m^3}$. It is important to note that there are two fundamental differences between the RED and the RBED previously introduced for semiconductors.\cite{Huntzinger2006, Mlayah2007, Large2010} First, the RED combines electric field and polarization modulation into a product homogeneous to an energy density while the RBED considers electronic wavefunctions to give an effective electronic density. Second, contrary to its semiconductor analog, the RED includes the vibrational component through $\delta_{m}\mathbf{P}_i(\mathbf{r})$.

  The RED can be divided into two terms:
\begin{eqnarray}\label{RED}
  \mathcal{U}_{\text{R},m}(\mathbf{r},\omega_{i},\mathbf{k}_i,\omega_{f},\mathbf{k}_f)= &  -\epsilon_0\delta_{m}\chi(\mathbf{r})\mathbf{E}_f(\mathbf{r})\cdot\mathbf{E}_i(\mathbf{r}) \nonumber \\
&  -\epsilon_0\chi(\mathbf{r})\mathbf{E}_f(\mathbf{r})\cdot\delta_{m}\mathbf{E}_i(\mathbf{r}),
\end{eqnarray}
where $\mathbf{E}_i(\mathbf{r})$ and $\mathbf{E}_f(\mathbf{r})$ are the classical local electric fields associated to the LSP before and after the emission of the acoustic vibrations, respectively. $\chi$ is the electric susceptibility of the metallic nanoparticle and is here defined using the Drude model:
\begin{equation}\label{chi}
\chi(\omega_{i},R)=\chi^\text{ib}(\omega_{i})-\frac{\omega_p^2}{\omega_{i}^2+i\omega_i\gamma(\omega_i,R)},
\end{equation}
where $\chi$ is dependent on the nanoparticle radius, $R$, through the size-corrected Drude damping $\gamma(\omega_{i},R) =\gamma_0+g_{\rm s}(\omega_i)v_{\rm F}/R$, where $v_{\rm F}$ is the Fermi velocity, $\gamma_0$ is the effective Drude damping, and $g_{\rm s}$ is obtained from a quantum treatment of the electron-surface scattering and is only weakly dependent on $\omega_i$.\cite{Apell1983,Hovel1993}

\subsubsection{Volume Mechanism}
The first term in Eq.~\ref{RED} describes the modulation of the electric susceptibility by the acoustic vibrations through $\delta_{m}\chi(\mathbf{r})$ and is called the deformation potential coupling mechanism or volume mechanism.\cite{Bachelier2004} A change in the nanoparticle volume results in a change in the electronic band structure \textit{via} the deformation potential and hence in the electric susceptibility. This particular mechanism has been the subject of a recent study by Saison-Francioso \textit{et al.}, which focused on the computational study of shape effect, electron density effect, and interband transition effects though the deformation potential mechanism.\cite{SaisonFrancioso2020} The effect of the electric susceptibility modulation on the intraband transitions is negligible in the visible range (where the LSP is located) because their excitation lies in the infrared range. Whereas, the interband transition modulation contributes to the volume mechanism as they occur in the UV-visible range, close to the LSP.\cite{Cottancin2006} The volume mechanism (first) term in Eq.~\ref{RED} reads
\begin{equation}\label{VolM}
\mathcal{U}_{\text{R},m}^\text{VM}(\mathbf{r})=-\epsilon_0\chi(\mathbf{r})\left[\frac{V_\text{DP}}{\hbar\omega_i-\hbar\omega_\text{ib}}\nabla\cdot\mathbf{u}_{m}(\mathbf{r})\right]\mathbf{E}_f(\mathbf{r})\cdot\mathbf{E}_i(\mathbf{r}),
\end{equation}
where $\hbar\omega_\text{ib}$ is the interband transition threshold, $V_\text{DP}$ is the deformation potential, and $\nabla\cdot\mathbf{u}_m$ is the divergence of the displacement field $\mathbf{u}_m(\mathbf{r})$ associated with the acoustic vibration mode $m$ and obtained from the elastodynamic calculations.

\subsubsection{Surface Mechanism}
The second term in Eq.~\ref{RED} corresponds to the surface orientation coupling mechanism.\cite{Bachelier2004} Contrary to the volume mechanism, the surface mechanism is the dominant contribution to the Raman scattering process in metals because of the strong plasmonic field localized at the nanoparticle surface.\cite{Bachelier2004, Large2009} As it is well known, the LSPs strongly depend on the nanoparticle shape. Therefore, the polarization induced by the LSP at the nanoparticle surface experiences a modulation by the acoustic vibrations through a change in the nanoparticle shape. A simple approximation to account for this surface orientation mechanism consists in a geometrical framework making use of the geometric factors in the nanoparticle polarizability tensor.\cite{Apell2022} Here, however, we rigorously describe the surface mechanism (second term in Eq.~\ref{RED}) in terms of the difference between the polarization fields of the final and initial LSP states. Then the second term in Eq.~\ref{RED} reads
\begin{equation}\label{SurM}
\mathcal{U}_{\text{R},m}^\text{SM}(\mathbf{r})=-\epsilon_0\chi(\mathbf{r})\mathbf{E}_f(\mathbf{r})\cdot\left[\mathbf{E}_f(\mathbf{r})-\mathbf{E}_i(\mathbf{r})\right],
\end{equation}
where we used $\mathbf{P}_{i(f)}(\mathbf{r})=\epsilon_0\chi(\mathbf{r})\mathbf{E}_{i(f)}(\mathbf{r})$.

\subsection{Raman Scattering Spectrum}
The Stokes Raman scattering spectrum of a metallic nanoparticle, derived from Fermi's golden rule for the acousto-plasmonic coupling (Eq.~\ref{Fermi_RED}), is expressed as
\begin{eqnarray}\label{Raman}
\displaystyle I_{\rm Raman} & = & \displaystyle \sum_m \frac{2\pi}{\hbar}\left| \int \mathcal{U}_{\text{R},m}(\mathbf{r}) dV \right|^2 \nonumber \\
& &\displaystyle \times \frac{\Gamma_m}{(\hbar\omega_{f}+\hbar\omega_{m}-\hbar\omega_{i})^2+(\Gamma_m/2)^2} \\ \nonumber 
& &\displaystyle \times \left[ \frac{1}{e^{(\hbar\omega_m/k_BT)}-1} +1 \right] 
\end{eqnarray}
where the summation runs for all the acoustic vibration modes. $\mathcal{U}_{\text{R},m}$, $\Gamma_m$, and $\omega_m$ are the RED, spectral linewidth, and eigenfrequencies, respectively, of the acoustic vibration mode $m$. The $\delta(\hbar\omega_{f}+\hbar\omega_{m}-\hbar\omega_{i})$ function is defined using a Lorentz distribution to account for the homogeneous broadening $\Gamma_m$. The term $\left[e^{(\hbar\omega_m/k_BT)}-1 \right]^{-1}$ is the Bose-Einstein population factor. Therefore, integrating the local Raman energy density yields the far-field low-frequency acoustic Raman scattering spectrum from the plasmonic nanoparticle.

\section{Computational Model and Methods}
\subsection{Computational Model}
\begin{figure}[!htbp]
\begin{center}
\includegraphics[width=\columnwidth,angle=0]{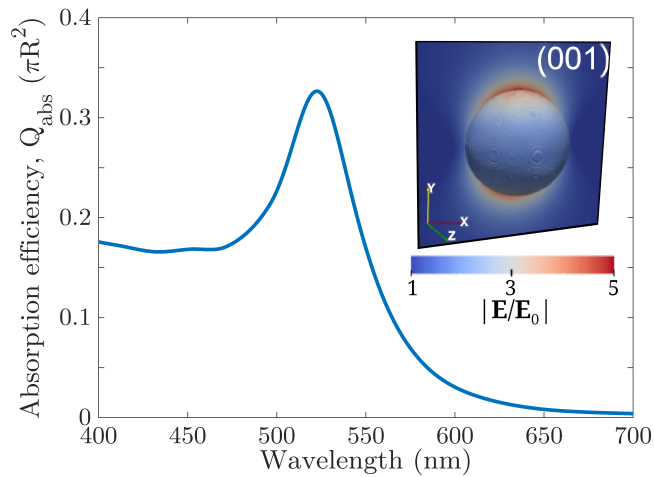}
\caption{Calculated absorption spectrum of a 5-nm AuNP in water showing the dipolar LSPR at 520 nm. Inset: 3D color map and (001) cross-sectional plane of the ENF at the LSPR.} \label{Fig1}
\end{center}
\end{figure}

To demonstrate and illustrate the concept of RED, we calculated the RED and computed the Raman spectrum of a free-standing Au spherical NP of radius $R = 2.5$ nm (Fig.~\ref{Fig1}a), immersed in water (refractive index $\mathit{n} = 1.333$), and optically exited at its LSPR ($\lambda_\text{LSPR}=520$ nm, Fig.~\ref{Fig1}). The incident optical excitation is a plane wave of unitary amplitude, polarized in \textit{y}-direction, and traveling along the $x$-direction. 

 In our model, we neglect any mechanical interaction between the NP and the surrounding aqueous medium.\cite{Voisin2002} However, it is important to note that the mechanical coupling between a more rigid environment and the nanoparticle may need to be taken into consideration. Furthermore, the heat generated by the NP under optical excitation has a negligible effect on the vibration frequencies and the LSPR, and will be neglected in our calculations.\cite{Ahmed2017}

\subsection{Elastodynamics: Resonant Ultrasound Spectroscopy (RUS) Method}
RUS is a method based on the measurement of the frequencies of the free vibrations of a solid object in order to determine the elastic tensor of the material. It has been successfully applied to systems ranging from small metallic clusters to macroscopic geological objects.\cite{Saviot2021} The vibrations are calculated using Rayleigh-Ritz variational approach. The displacement vector field $\mathbf{u}(\mathbf{r})$ associated with each acoustic vibration mode is calculated assuming an elastic continuous medium\cite{Combe2009} and using the $xyz$ algorithm proposed by Visscher \textit{et al.} to solve the Navier-Stokes hydrodynamic equations.\cite{Visscher1991} 
  RUS uses the longitudinal and transverse sound velocities in the material and expands the displacement vectors onto a basis $\Phi_{\zeta}=x^iy^jz^k$, where $\zeta=(i,j,k)$ is the function label. Here, we used $0 \leq i + j + k \leq 20$ in order to achieve a good numerical convergence.\cite{Saviot2021} The Rayleigh-Ritz approach takes the
dynamic problem into a generalized eigenvalue problem where all the eigenmodes obtained are orthonormalized.\cite{Saviot2021, Murray2004} The values used for the Au density ($\rho$), longitudinal and transverse sound velocities ($v_{\rm L}$ and $v_{\rm T}$, respectively), and Poisson's ratio ($\nu$) are provided in Table~\ref{GoldParam}.

\subsection{Electrodynamics: Discrete Dipole Approximation (DDA)}
Once the vibration modes have been calculated in RUS, we use DDA (DDSCAT v7.3 package)\cite{Draine1994} to calculate the local electric field at the LSPR wavelength for the particle at rest (initial state) and for the nanoparticle deformed by a given acoustic vibration mode. In all the DDA simulations, we used an interdipole distance $d = 0.03$ nm to discretize the nanoparticle and achieve numerical convergence; the total number of dipoles within the nanoparticle is around $3\times 10^6$. The physical parameters used for gold in Eq.~\ref{chi} are provided in Table~\ref{GoldParam}. 

\begin{table*}
\begin{center}
\captionof{table}{Gold parameters used for the RED and Raman spectrum calculations.}
\label{GoldParam}
\begin{tabular}{ c c c | c c c}
\multicolumn{3}{c|}{Electrodynamics Simulation (DDA)} & \multicolumn{3}{c}{Elastodynamics Simulation (RUS)} \\ 
 Parameter & Value & Reference & Parameter & Value & Reference \\
 \hline\hline 
 $\chi^\text{ib}$ & J\&C database & ~\citenum{Johnson1972}~ & $\rho$ & 19,700 kg/m$^{3}$ & ~\citenum{Voisin2002}~ \\  
 $\hbar\omega_{\rm p}$ & 9.01 eV & ~\citenum{Derkachova2016}~ & $v_{\rm L}$ & 3240 m/s & ~\citenum{Voisin2002}~ \\
 $\gamma_0$ & 0.07 eV & ~\citenum{Derkachova2016}~ & $v_{\rm T}$ & 1200 m/s & ~\citenum{Voisin2002}~  \\
 $g_{\rm s}v_{\rm F}$ & 0.915 eV\,nm & ~\citenum{Derkachova2016}~  & $\nu$ & 0.42 & ~\citenum{Voisin2002}~ \\
 $V_\text{DP}$ & -0.8 eV & ~\citenum{Szczepanek1974}~  &  &  & \\   
 $\hbar\omega_\text{ib}$ & 2.4 eV & ~\citenum{Derkachova2016}~ & & & \\
 \hline
\end{tabular}
\end{center}
\end{table*}

\section{Results and Discussions}
\subsection{Acoustic Vibrations}
\begin{figure}[!htbp]
\begin{center}
\includegraphics[width=\columnwidth,angle=0]{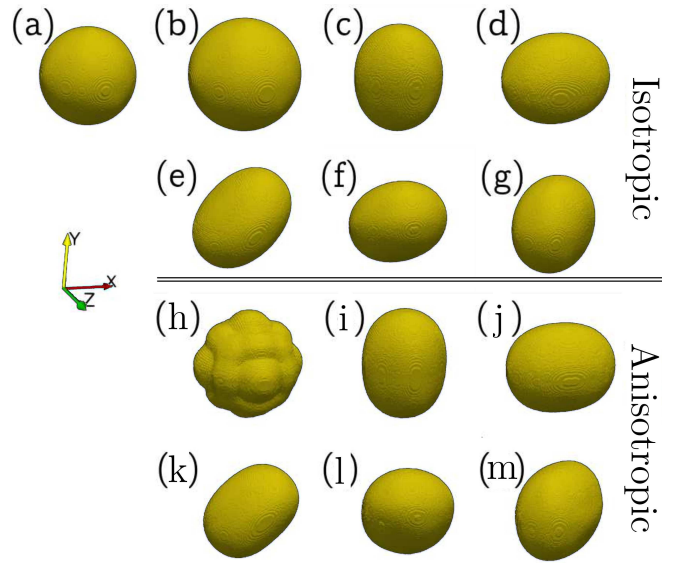}
\caption{(a) AuNP at rest. Top: NP deformed by the isotropic breathing $S^1_{00}$ ($d_0$) mode (b) and the fivefold degenerated quadrupole $S^1_{2m}$ ($d_1-d_5$) modes (c-g). Bottom: NP deformed by the anisotropic breathing $\rm{A_{1g}}$ mode (h), the quadrupole $\rm{E_g}$ [$\rm{E_{g(a)}}$ and $\rm{E_{g(b)}}$] modes (i-j), and the quadrupole $\rm{T_{2g}}$ [$\rm{T_{2g(a)}}$, $\rm{T_{2g(b)}}$, and $\rm{T_{2g(c)}}$] modes (k-m).} \label{Fig2}
\end{center}
\end{figure}

To compute the Raman scattering spectrum of a AuNP undergoing acoustic vibrations and the associated RED, we first calculate the surface displacement of a free-standing, homogeneous, isotropic, and elastic Au nanosphere induced by the acoustic vibration modes. In Lamb's original paper, the acoustic vibration modes were classified into torsional and spheroidal modes.\cite{Lamb1881} Torsional modes do not induce any change in the materials density, which implies that the divergence of the displacement is $\nabla\cdot\mathbf{u}=0$, thus leading to $\mathcal{U}_\text{R}^\text{VM}=0$. Moreover, because of the absence of shape change induced by such modes, we also have $\mathcal{U}_\text{R}^\text{SM}=0$. Therefore, because they do not contribute to either the surface or the volume mechanism, these torsional modes are Raman inactive and will thus not be considered in this work. On the other hand, spheroidal modes induce changes in the NP shape and/or volume. These modes are labeled $S^n_{\ell m}$; where the integers $n$, $\ell$, and $m$ denote the harmonic ($n=1$ being the fundamental), the angular momentum number, and its $z$-component azimuthal number, respectively. The vibrational density of states is discrete and the mode eigenfrequencies are given by $\omega_{\ell,n}[{\rm cm^{-1}}]=\xi_{\ell,n}v_{\rm L}/2R$, where $\xi_{\ell,n}$ is a mode-dependent coefficient.\cite{Voisin2002} Figures~\ref{Fig2} and S1 (Supplementary Materials)\cite{SupMat} show the surface displacements associated with the fundamental breathing mode ($S^1_{00}$; Fig.~\ref{Fig2}b) and the fivefold degenerated fundamental quadrupole modes ($S^1_{2m}$ with $m=0,\pm1,\pm2$; Fig.~\ref{Fig2}c-g). 

  Because gold exhibits a strong elastic anisotropy, we have also calculated the surface displacements associated with the six equivalent breathing and quadrupolar vibration modes from an anisotropic Au nanosphere. The irreducible representations of the anisotropic vibration modes have been determined based on the $\rm{O_h}$ point group character table.\cite{Saviot2009} As a result of the elastic anisotropy, the spheroidal breathing modes transform into an $\rm{A_{1g}}$ vibration (Fig.~\ref{Fig2}h). Similarly, the anisotropy induces a partial degeneracy lift of the fivefold spheroidal quadrupolar mode which splits into two $\rm{E_{g}}$ (Fig.~\ref{Fig2}i-j) and three $\rm{T_{2g}}$ (Fig.~\ref{Fig2}k-m) degenerated vibrations; the latter can be projected onto spheroidal Lamb modes.\cite{Saviot2009}

\begin{table*}
\begin{center}
\captionof{table}{Eigenfrequencies of the isotropic and anisotropic vibration modes calculated using RUS and compared to analytic results from Lamb theory.}
\label{eigenfrequencies}
\begin{tabular}{cc|cc|cc}
\multicolumn{2}{c|}{Lamb theory} & \multicolumn{2}{c|}{Isotropic (RUS)} & \multicolumn{2}{c}{Anisotropic (RUS)} \\
\hline \hline
Mode           & Frequency        & Mode      & Frequency        & Mode    & Frequency \\
$S^n_{\ell m}$ & ($\rm{cm^{-1}}$) &           & ($\rm{cm^{-1}}$) &	     & ($\rm{cm^{-1}}$) \\
\hline
$S^1_{00}$     &  20.8    & $d_0$ (Fig.~\ref{Fig2}b) & 20.8  & $\rm{A_{1g}}$ (Fig.~\ref{Fig2}h)  & 20.7 \\
$S^1_{20}$     &  7.1   & $d_1$ (Fig.~\ref{Fig2}c) & 7.1   & $\rm{E_{g(a)}}$ (Fig.~\ref{Fig2}i)  & 5.0\\
$S^1_{2\pm1}$  &  7.1   & $d_2$ (Fig.~\ref{Fig2}d) & 7.1   & $\rm{E_{g(b)}}$ (Fig.~\ref{Fig2}j)  & 5.0\\
$S^1_{2\mp1}$  &  7.1   & $d_3$ (Fig.~\ref{Fig2}e) & 7.1   & $\rm{T_{2g(a)}}$ (Fig.~\ref{Fig2}k) & 8.0 \\
$S^1_{2\pm2}$  &  7.1   & $d_4$ (Fig.~\ref{Fig2}f) & 7.1   & $\rm{T_{2g(b)}}$ (Fig.~\ref{Fig2}l) & 8.0 \\
$S^1_{2\mp2}$  &  7.1   & $d_5$ (Fig.~\ref{Fig2}g) & 7.1   & $\rm{T_{2g(c)}}$ (Fig.~\ref{Fig2}m) & 8.0 \\
\hline
\end{tabular}
\end{center}
\end{table*}

The eigenfrequencies and the labels we use for each vibration mode are presented in Table~\ref{eigenfrequencies}. Because of the weak mechanical coupling between the nanoparticle and the surrounding environment,\cite{Voisin2002} the nanoparticle can be considered as free standing. Therefore, we also compare the eigenfrequencies obtained using RUS to analytical results from Lamb theory.\cite{Lamb1881}
It is clear from Fig.~\ref{Fig2} that the directions of maximum deformation are mode-dependent. The breathing mode ($d_0$) is isotropically deformed in all directions due to its pure radial nature, thus preserving the spherical shape throughout the vibration period (Fig.~\ref{Fig2}b). On the other hand, the anisotropic breathing mode ($\rm{A_{1g}}$) has maximum deformations in the [100] directions, \textit{i.e.}, along the $x$-, $y$-, and $z$-directions, thus breaking the spherical symmetry (Fig.~\ref{Fig2}h). For the sake of presentation, we have chosen a cross-section plane of the NP which contains the direction of maximum deformation for each vibration mode. Figure~S1 of the Supplementary Materials shows the aforementioned cross-section planes of the surface displacements of a AuNP for each isotropic (blue line) and anisotropic (red line) vibration modes relative to the NP at rest (black line).\cite{SupMat} These planes of maximum deformation are also chosen to contain the polarization direction ($y$-direction) of the optical excitation used in the electrodynamic calculations. Although more pronounced for the breathing modes, the difference on the surface deformations between isotropic and anisotropic vibration modes are clearly visible (Fig. S1, Supplementary Materials\cite{SupMat}).

\subsection{Plasmonic Near-Field}
In order to compute the RED, we have calculated the electric near-field (ENF) of the AuNP at rest excited at its LSPR (Fig.~\ref{Fig1}) with a plane wave propagating in the $x$-direction and polarized along the $y$-direction (\textit{i.e.}, initial state). The inset in Fig.~\ref{Fig1} shows the ENF spatial distribution at the surface of the NP and in the (001) plane ($xy$-plane); it clearly shows the dipolar nature of the LSP and a maximum electric field enhancement $|{\bf E}/{\bf E}_0|\approx 5$.

\begin{figure}[!htbp]
\begin{center}
\includegraphics[width=\columnwidth,angle=0]{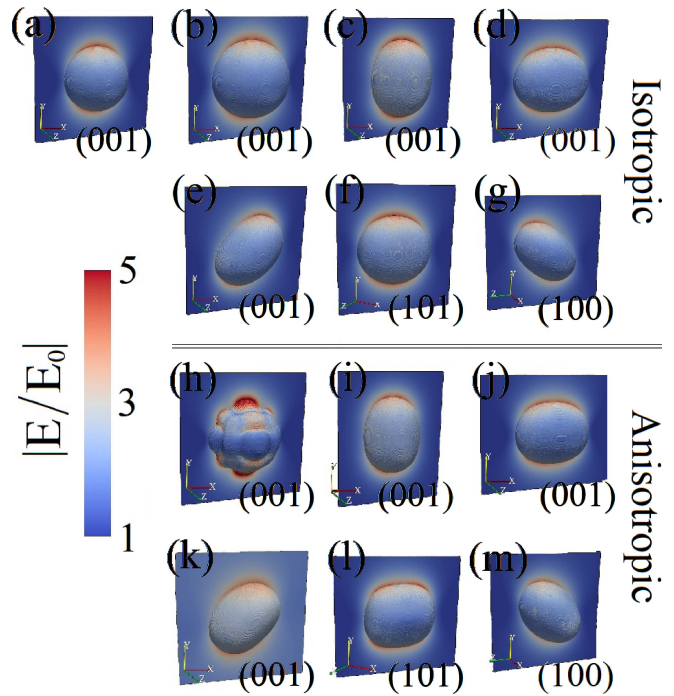}
\caption{2D/3D spatial distributions of the ENF enhancement induced by the dipolar LSPR and modulated by the isotropic (top) and anisotropic (bottom) vibration modes. The cross-sections are taken where the maximum deformation is located; the corresponding cross-section planes are indicated in each panel. (a) NP at rest, (b) isotropic breathing $S^1_{00}$ ($d_0$) mode, (c-g) fivefold degenerated isotropic quadrupoles $S^1_{2m}$ ($d_1$-$d_5$), (h) anisotropic breathing $\rm{A_{1g}}$ mode, (i,j) anisotropic quadrupole $\rm{E_g}$ [$\rm{E_{g(a)}}$ and $\rm{E_{g(b)}}$] modes, and (k-m) anisotropic quadrupole $\rm{T_{2g}}$ [$\rm{T_{2g(a)}}$, $\rm{T_{2g(b)}}$, and $\rm{T_{2g(c)}}$] modes.} \label{Fig3}
\end{center}
\end{figure}

  Figure~\ref{Fig3} shows the ENF spatial distributions on the surface of the NPs deformed by the vibration modes as well as in the cross-sectional planes of maximum deformation. As discussed earlier, this figure shows how the ENF is distinctively and spatially modulated by each individual vibration mode. It is important to notice that the amplitude of the modulation depends on the mode. This, will ultimately reflect on the Raman efficiency of the mode as we will discuss further. Dynamic ENF modulations for all the acoustic vibration modes are shown as ENF-ISO.gif and ENF-ANISO.gif (Supplementary Materials).\cite{SupMat} 

The ENF distribution of the AuNP deformed by the isotropic breathing mode (Fig.~\ref{Fig3}b) does not appear to change in comparison with that of the NP at rest (Fig.~\ref{Fig3}a). This absence of modulation from the isotropic breathing mode is due to (i) the conservation of the spherical shape and (ii) the size change smaller than the optical wavelength and negligible retardation effects. Consequently, according to Eq.~\ref{SurM}, the surface mechanism contribution to the RED will be negligible. On the other hand, the anisotropic breathing mode (Fig.~\ref{Fig3}h) appears to have the largest ENF enhancement due to the strongest localization of surfaces charges at the NP poles (along the $y$-direction). The same occurs for the $d_1$ and $\rm{E_{g(a)}}$ vibration modes (Fig.~\ref{Fig3}c,i), where the deformation of the NP is also along the applied electric field, yielding a little more enhancement compared to the NP at rest. For the other modes, the ENF is imperceptibly modulated because the surface displacements of the vibrating NP occur in directions other than that of the incident electric field polarization. Figure~S2 of the Supplementary Materials also shows the ENF enhancement spatial distributions in the planes of maximum deformation, where the $\Delta$-axis refers to the direction of the specific plane (\textit{e.g.}, $\Delta$ is equivalent to $x$ for the (001) plane).\cite{SupMat}

\subsection{Raman Energy Density (RED)}
We have calculated the partial RED associated with the surface (Eq. ~\ref{SurM}) and volume (Eq. ~\ref{VolM}) mechanisms as well as the total (Eq. ~\ref{RED}) RED (in $\rm{neV/nm^3}$) for each isotropic and anisotropic vibration mode. It should be noted that the region between the surface of the nanoparticle at rest and the deformed surface was not considered and set to zero due to the lack of physical realism. Indeed, the surface location cannot be defined with a precision better than the interdipole spacing in the DDA simulations.

\begin{figure}[!htbp]
\begin{center}
\includegraphics[width=\columnwidth,angle=0]{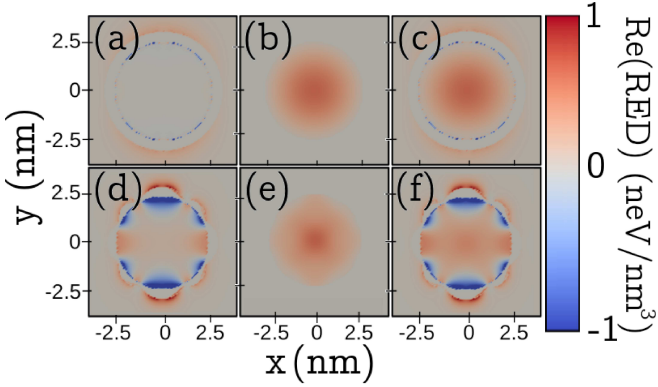}
\caption{Real part of the surface $\mathcal{U}_\text{R}^\text{SM}$ (a,d), volume $\mathcal{U}_\text{R}^\text{VM}$ (b,e), surface+volume $\mathcal{U}_R$ (c,f) components of the RED associated with the isotropic (a-c) and the anisotropic (d-f) breathing modes.} \label{Fig4}
\end{center}
\end{figure}

  Figure~\ref{Fig4} displays the spatial distributions of the real part of $\mathcal{U}_\text{R}^\text{SM}$, $\mathcal{U}_\text{R}^\text{VM}$, and $\mathcal{U}_\text{R}$ for the isotropic (Fig.~\ref{Fig4}a-c) and anisotropic (Fig.~\ref{Fig4}d-f) breathing modes. These maps show the spatial extent of the coupling between the acoustic vibrations and the dipolar LSP yielding inelastic light scattering in and around the NP. They represent the acousto-plasmonic local source of the Raman scattering measured in the far-field. The partial RED associated with the deformation of the NP surface ($\mathcal{U}_\text{R}^\text{SM}$) by the $d_0$ mode is localized only on the NP surface; there is no acousto-plasmonic coupling inside the NP. When anisotropy is included ($\rm{A_{1g}}$ mode), $\mathcal{U}_\text{R}^\text{SM}$ is enhanced thanks to the localization of the LSP at the deformed surface and to the acousto-plasmonic coupling occurring inside the NP. On the other hand, $\mathcal{U}_\text{R}^\text{VM}$ shows a radially increasing acousto-plasmonic coupling from the surface to the center of the NP, which is mainly induced by the larger expansion (or compression) inside the NP than underneath the surface (Fig. S3, Supplementary Materials\cite{SupMat}).\cite{Ahmed2017} The volume contribution to the RED, as seen in panels b and e, arises from the local variations of the electric susceptibility induced by the vibration modes through the deformation potential mechanism (Eq.~\ref{VolM}). This result is in good agreement with the local volume variations of the dielectric permittivity produced by the NP vibration modes and which have been recently predicted by Saison-Francioso \textit{et al.}\cite{SaisonFrancioso2020}

The total RED ($\mathcal{U}_\text{R}$, Fig.~\ref{Fig4}c,f) clearly appears to be more spatially localized in the case of the $\rm{A_{1g}}$ mode compared to the $d_0$ mode. These RED hot-spots correspond to regions where the NP simultaneously exhibits a strong surface deformation and a strong induced ENF. Nevertheless, it is important to notice that the Raman activity of each acoustic vibration mode is obtained by spatially integrating the RED (Eq.~\ref{Raman}). As a result of this spatial integration, positive and negative quantities may cancel each other out, thus leading to a decrease or to a complete cancellation of the Raman response, despite the presence of RED hot-spots. Based on this observation, the $d_0$ mode is expected to lead to a larger Raman efficiency than the $\rm{A_{1g}}$ mode, because the former has no negative contribution to the RED.

\begin{figure}[h]
\begin{center}
\includegraphics[width=\columnwidth,angle=0]{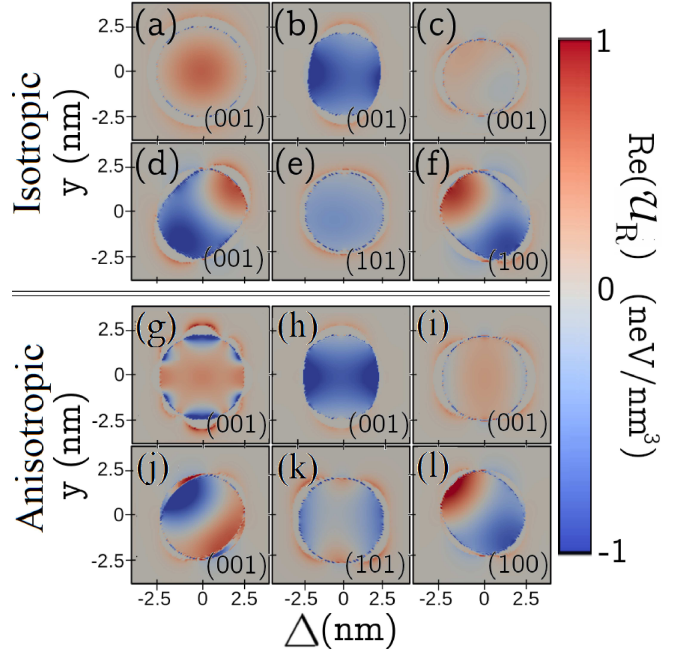}
\caption{Real part of the total RED, $\mathcal{U}_R$, associated with the isotropic vibration modes [$d_0$ (a), $d_1$ (b), $d_2$ (c), $d_3$ (d), $d_4$ (e), $d_5$ (f)] and anisotropic vibration modes [$\rm{A_{1g}}$ (g), $\rm{E_{g(a)}}$ (h), $\rm{E_{g(b)}}$ (i), $\rm{T_{2g(a)}}$ (j), $\rm{T_{2g(b)}}$ (k), $\rm{T_{2g(c)}}$ (l)].} \label{Fig5}
\end{center}
\end{figure}

The same analysis applies for the other isotropic and anisotropic modes. Figure~\ref{Fig5} shows the total RED, $\mathcal{U}_\text{R}$, calculated for each isotropic and anisotropic mode. The two RED components $\mathcal{U}_\text{R}^\text{SM}$ and $\mathcal{U}_\text{R}^\text{VM}$ of the same modes are presented in Fig.~S4 (Supplementary Materials).\cite{SupMat} For most of the fivefold degenerated quadrupolar modes, the contribution of the $\mathcal{U}_\text{R}^\text{SM}$ in the total RED is larger than $\mathcal{U}_\text{R}^\text{VM}$ in both the isotropic and anisotropic cases. Nevertheless, as discussed above, because of cancellation of the negative and positive contributions to the RED, one expects a lower Raman efficiency for $\rm{A_{1g}}$. As for example, it could be expected that the contribution of the $d_3$, $d_4$, and $d_5$ modes (Fig.~\ref{Fig5}d-f) to the Raman scattering will be low, also for $\rm{T_{2g}}$ modes (Fig.~\ref{Fig5}j-l). On the other hand, it is also expected that the contribution of the $d_1$ and $\rm{E_{g(a)}}$ modes will be large owing to their high RED intensity inside the NP, mostly from the $\mathcal{U}_\text{R}^\text{SM}$ contribution (center panels of Fig. S4b, Supplementary Materials\cite{SupMat}). Although $d_2$ and $\rm{E_{g(b)}}$ modes have positive values (Fig.~\ref{Fig5}c and~\ref{Fig5}i, respectively), their contribution to the Raman scattering is likely to be low due to their weak RED values.

\subsection{Acoustic Raman Spectra}
\begin{figure}[!htbp]
\begin{center}
\includegraphics[width=\columnwidth,angle=0]{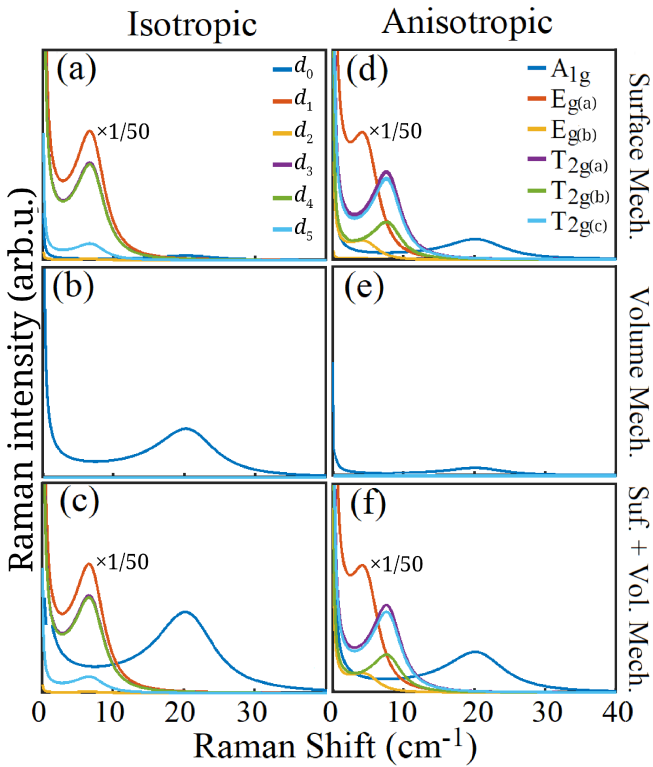}
\caption{Raman spectra calculated using Eq.~\ref{Raman} for the surface (a,d), volume (b,e), and both (c,f) mechanisms of the isotropic (a-c) and anisotropic (d-f) vibration modes} \label{Fig6}
\end{center}
\end{figure}

Figure~\ref{Fig6}a-c displays the Raman spectra calculated for each isotropic vibration mode using Eq.~\ref{Raman} and considering the surface and volume mechanisms as well as the sum of the two mechanisms.

  The $d_1$ mode gives the strongest Raman response because its maximum displacement is parallel to the incident electric field polarization. On the other hand, $d_0$ and $d_2$ have very weak contributions through the surface mechanism because the former leads only to small changes of the NP shape and the latter exhibits the maximum deformation in the direction perpendicular to the incident polarization. The only mode that leads to a strong contribution to the Raman efficiency through the volume mechanism is $d_0$. When the NP undergoes a $d_0$ mode, it experiences a global expansion (or compression) unlike the other modes, which on the contrary experience simultaneous compression and expansion at different locations on the NP surface (center panel of Fig.~S4, Supplementary Materials\cite{SupMat}), and thus resulting in the cancellation of the Raman efficiency.

On the other hand, the Raman spectra calculated for each anisotropic vibration mode and considering the surface and volume mechanisms as well as the sum of the two mechanisms are presented in Fig.~\ref{Fig6}d-f, respectively. Similarly to the isotropic case, the quadrupolar modes produce a strong Raman intensity through the surface mechanism while the breathing mode contributes through the volume mechanism. However, a few differences can be noticed: the $\rm{T_{2g}}$ modes Raman peaks are shifted to higher energy and are less intense than their isotropic counterparts $d_3$, $d_4$, and $d_6$ due to the anisotropy in sound velocities.

\begin{figure}[!htbp]
\begin{center}
\includegraphics[width=\columnwidth,angle=0]{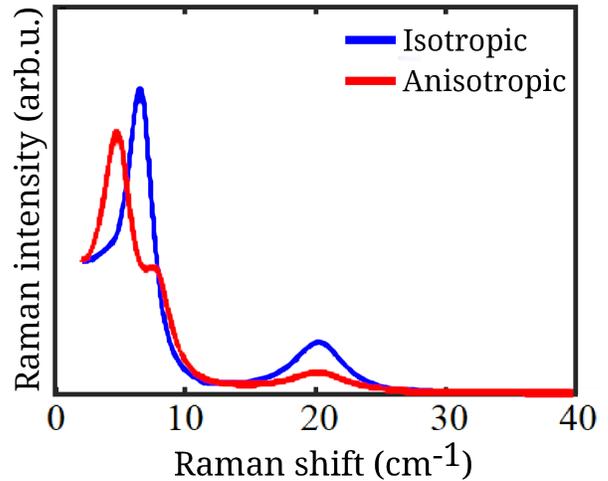}
\caption{Total Raman spectra calculated using Eq.~\ref{Raman} for the isotropic and anisotropic modes.} \label{Fig7}
\end{center}
\end{figure}

As a consequence, the Raman spectrum of an isotropic AuNP has principally two peaks, which correspond to the acousto-plasmonic coupling of the fivefold degenerated quadrupolar vibration modes \textit{via} surface mechanism and of the breathing mode \textit{via} the volume mechanism (Fig.~\ref{Fig7}, blue). The Raman spectrum of an anisotropic AuNP has three peaks. The two lower-frequency peaks are due to the quadrupolar $\rm{E_g}$ and $\rm{T_{2g}}$ modes coupled to the LSP \textit{via} the surface mechanism while the higher energy peak is due to the breathing $\rm{A_{1g}}$ mode which is Raman active \textit{via} the volume coupling mechanism (Fig.~\ref{Fig7}, red).

  As previously discussed, the positive and negative contributions of the RED associated with the anisotropic breathing mode cancel each other out, thus leading to a decrease in the Raman intensity and resulting in the $d_0$ mode being more intense than the $\rm{A_{1g}}$ mode.
As for the quadrupolar modes, it is important to note that we recover the intensity relationship between the isotropic and anisotropic modes. Indeed, the sum of the intensities of the isotropic quadupolar modes (\textit{i.e.}, $\sum_{k=1}^5 d_k$) is equal to the sum of the intensities of the anisotropic quadrupolar modes (\textit{i.e.}, $\sum_{k=a}^c {\rm E}_{{\rm g}(k)}+{\rm T}_{{\rm 2g}(k)}$), with a relative error of $1.5\times 10^{-3}$.

  In both the isotropic and anisotropic cases, the quadrupolar $d_1$ and $\rm{E_{g(a)}}$ modes are the most intense as previously discussed. These results are in excellent quantitative agreement with several experiments performed on small AuNPs.\cite{Saviot2012, Tripathy2011} In particular, the measured Raman intensity of the $\rm{T_{2g}}$ modes is weaker as compared to the $\rm{E_g}$ modes, in agreement with our theoretical calculations (Fig.~\ref{Fig7}, red spectrum). Once again, the difference in intensity between these two anisotropic quadrupolar mode symmetries lies in their RED components. Indeed, as shown in Fig.~\ref{Fig5}, the real parts of the RED associated with the $\rm{T_{2g}}$ modes (panels j-l) exhibit positive and negative contributions, which tend to cancel out in Eq.~\ref{Fermi_RED}. On the other hand, the real parts of the RED associated with the $\rm{E_{g}}$ modes (panels h-i) are predominantly either positive or negative.

\subsection{Raman Selection Rules}
\begin{figure}[!htbp]
\begin{center}
\includegraphics[width=\columnwidth,angle=0]{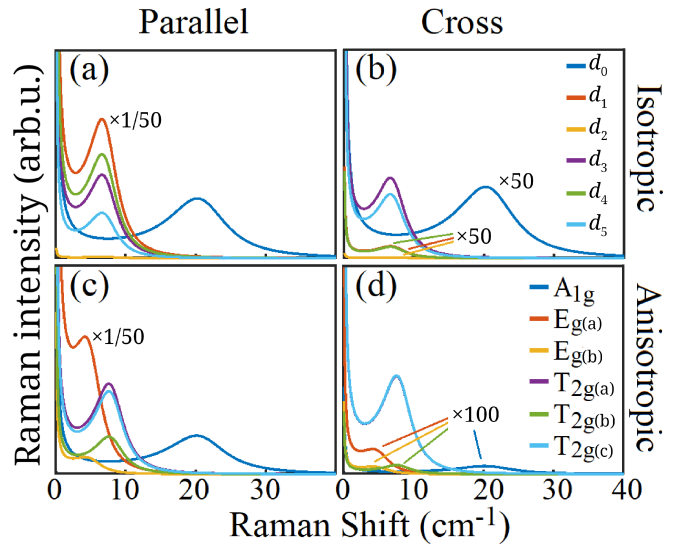}
\caption{Raman spectra calculated using Eq.~\ref{Raman} for the parallel (a,c) and cross (b,d) configurations of the isotropic (a,b) and anisotropic (c,d) modes.} \label{Fig8}
\end{center}
\end{figure}

Finally, we compute the Raman spectra using Eq.~\ref{Raman} for various incident and scattered polarization configurations to determine the Raman selection rules. We demonstrate the contribution of each mode in the situation where an analyzer (polarization filter) parallel or perpendicular to the incident electric field is used before the Raman detector (Fig.~\ref{Fig8}). First, we see that all mode Raman intensities have slightly decreased when a parallel analyzer is used in both the isotropic and anisotropic cases. Furthermore, when using a perpendicular analyzer (cross configuration), the breathing mode contribution becomes negligible and only a few quadrupolar contributions persist. It is well know from experiments\cite{Mlayah2007} that the breathing mode of a spherical nanoparticle is totally polarized, whereas the quadrupolar modes are only partially polarized. This can easily be understood when looking at the RED components (Fig.~S4, Supplementary Materials\cite{SupMat}). Due to its spherical symmetry, the modulation produced by the isotropic breathing mode, through the surface mechanism, has no component perpendicular to the incident polarization (top panel of Fig.~S4a, Supplementary Materials\cite{SupMat}). However, it is important to notice that the breathing mode does not completely vanish, due to the weaker, yet contributing, volume component of the RED (center panel of Fig.~S4a, Supplementary Materials\cite{SupMat}). 

Conversely, the quadrupolar modes induce a RED with components along the incident field direction (Fig.~S4b-f, Supplementary Materials\cite{SupMat}). Their individual symmetries, which is given by relative orientation of the major deformation axis with respect to the incident polarization, dictates their relative Raman intensities (Figs~\ref{Fig8} and S4, Supplementary Materials\cite{SupMat}). These observations are in excellent agreement with the theoretical work by Duval,\cite{Duval1992} who showed that for the Raman transition that excites a spherical breathing mode, the polarization of incident and scattered fields are parallel only, while they can be parallel or perpendicular in the case of the quadrupolar modes. Our theoretical approach also explains the selections rules observed experimental by Tripathy \textit{et al.} in the framework of acoustic SERS,\cite{Tripathy2011} Interestingly, we can note that the experiments conducted by Tripathy \textit{et al.} revealed a large number of harmonics for the various modes. While we limited our work to the fundamental modes ($n$=1) only, our approach can easily include harmonics to describe the experimental observations. For the harmonics ($n>$1), two important factors will impact the Raman intensity. First, the inner deformation associated with the presence of displacement nodes within the volume will directly impact the volume component of the RED. Second, the lower surface displacement amplitudes will result in weaker modulation of the ENF at the nanoparticle surface, thus directly impacting the surface component of the RED. Both of these effects lead to a decrease of the Raman efficiency with increasing overtones, making them challenging to observe.\cite{Bachelier2004, Saviot2012} While an atomistic description of the nanoparticle vibration eigenmodes is traditionally required,\cite{Combe2009, Saviot2012} our RED framework provides a new route to fully describe such high order harmonics in the acoustic Raman spectra.

\section{Conclusions}
We introduced the Raman energy density (RED) as a new theoretical tool for the interpretation of resonant Raman scattering mediated by LSPs in metallic nanostructures. Analogous to the LDOS, it represents the electromagnetic energy density, which is excited by the Raman probe and modulated by the acoustic vibrations of the nanostructure. It is a local quantity that can be mapped in the near-field region to provide a clear picture of the acousto-plasmonic interaction, which gives rise to the inelastic light scattering measurable in the far-field. It allowed us to calculate for the first time the Raman selection rules taking into account the surrounding medium and the modulation of the polarization in this medium. We showed that the RED can accurately predict the Raman-active modes and the ones forbidden by the Raman selection rules for both isotropic and anisotropic modes, in excellent quantitative agreement with other theoretical frameworks\cite{Duval1992,Saviot2012} and experiments.\cite{Tripathy2011,Saviot2012} The RED provides significant insight about the origin of the plasmon-vibration coupling (volume \textit{vs.} surface mechanisms). 

  It is worth noting that the RED framework goes well beyond plasmonic nanoparticles and can be applied to all type of nanomaterials including semiconductors and dielectrics. For instance, dielectric nanoparticles sustain Mie resonances, which induce local electric fields. In such case, similar electrodynamic simulations can be performed and coupled to the elastodynamic calculations to compute the resulting acoustic Raman scattering spectra and associated RED. Semiconductor quantum dots (QD) can also be described using this approach. While the local electric field induced by excitons is much weaker, leading to a small surface contribution, the volume deformation is expected to strongly modulate the QD electronic density of state; thus resulting in a strong volume contribution. Another aspect that can also be integrated into this framework concerns the mechanical properties of the surrounding medium, which affect the nanoparticles vibrational properties.\cite{Murray2004} Our approach can capture and provide insight into the mechanical and electromagnetic coupling through the RED and computed Raman spectra.

  The RED constitutes a novel and unique tool for the exploration of physical effects such as acoustic SERS,\cite{Tripathy2011} vibrational transfer,\cite{Girard2016} vibrational hybridization,\cite{Girard2018} sensing,\cite{Tripathy2011} photothermal effects and heat dissipation,\cite{Cunha2020} and other optomechanical phenomena between neighboring NPs, substrates, and surrounding medium.\cite{Deacon2017} It constitutes a step towards the realization of nanophononic platforms capable of probing simultaneously electronic, thermal, and phononic states.

\section*{Acknowledgments}
We thank Dr. Lucien Saviot from the Laboratoire Interdisciplinaire Carnot de Bourgogne at the Universit\'{e} de Bourgogne for the valuable and constructive discussions and for providing the RUS codes used for the calculation of the acoustic vibrations.
This work received financial support from the Army Research Office (ARO) under Grant Number W911NF-18-1-0439, the Office of Naval Research (ONR) under Grant Number N00014-21-1-2729, and the NanoX Grant ANR-17-EURE-0009 in the framework of the "Programme des Investissements d’Avenir". It received computational support from UTSA’s HPC cluster Shamu, operated by Tech Solutions.


%

\end{document}